\documentclass[conference, 9.96pt, a4paper]{IEEEtran}
\usepackage{acronym}
\acrodef{3GPP}{3rd generation partnership project}
\acrodef{3D}{three-dimensional}
\acrodef{2D}{two-dimensional}
\acrodef{5G}{fifth generation}
 \acrodef{AGV}{automatic guided vehicle}
 \acrodef{ADMM}{alternating direction method of multipliers}
\acrodef{BS}{base station}
\acrodef{CDF}{cumulative distribution function}
\acrodef{CSI}{channel state information}
\acrodef{CQI}{channel quality indicator}
\acrodef{CNN}{convolutional neural network}
\acrodef{CSP}{communications service provider}
\acrodef{DL}{deep learning}
\acrodef{DNN}{deep neural network}
\acrodef{DQN}{deep Q-network}
\acrodef{DRL}{deep reinforcement learning}
\acrodef{DDPG}{Deep Deterministic Policy Gradient}
\acrodef{XR}{extended reality}
\acrodef{FDD}{frequency division duplex}
\acrodef{FDMA}{frequency division multiple access}
\acrodef{GP}{Gaussian process}
\acrodef{GD}{Gradient Descent}
\acrodef{GUI}{graphical user interface}
\acrodef{HetNet}{heterogeneous network}
\acrodef{IoT}{internet of things}
\acrodef{IDLA}{integrated deep learning and Lagrangian method}
\acrodef{KL}{Kullback-Leibler}
\acrodef{KKT}{Karush–Kuhn–Tucker}
\acrodef{KPI}{key performance indicator}
\acrodef{LTE}{long term evolution}
\acrodef{M2M}{machine-to-machine}
\acrodef{MAC}{media access control}
\acrodef{MAE}{mean absolute error}
\acrodef{MAPE}{mean absolute percentage error}
\acrodef{MARL}{multi-agent reinforcement learning}
\acrodef{MADRL}{multi-agent deep reinforcement learning}
\acrodef{MDP}{Markov decision process}
\acrodef{ML}{machine learning}
\acrodef{MMDP}{multi-agent Markov Decision Process}
\acrodef{MIMO}{multiple-input and multiple-output}
\acrodef{mmWave}{millimeter wave}
\acrodef{MLP}{multi-layer perceptron}
\acrodef{MILP}{mixed integer linear programming}

\acrodef{NSM}{Network Slicing Management}
\acrodef{OAM}[O\&M]{Network Operation and Maintenance}
\acrodef{O-RAN}{Open Radio Access Network}
\acrodef{OFDM}{orthogonal frequency division multiplexing}
\acrodef{PASA}{production-aware slicing resource allocation}
\acrodef{PDF}{probability density function}
\acrodef{PF}{proportional fairness}
\acrodef{PPO}{proximal policy optimization}
\acrodef{PHY}{physical layer}
\acrodef{PSD}{power spectral density}
\acrodef{PRB}{physical resource block}
\acrodef{PRBs}{physical resource blocks}
\acrodef{QoE}{quality of experience}
\acrodef{QoS}{quality of service}
\acrodef{RAN}{radio access network}
\acrodef{RB}{resource block}
\acrodef{RL}{reinforcement learning}
\acrodef{RR}{round robin}
\acrodef{RRM}{radio resource management}
\acrodef{RU}{resource unit}
\acrodef{RX}{receiver}
\acrodef{SDN}{software defined network}
\acrodef{SNR}{signal-to-noise ratio}
\acrodef{SINR}{signal-to-interference-plus-noise ratio}
\acrodef{SIR}{signal-to-interference ratio}
\acrodef{SLA}{service level agreement}
\acrodef{SONs}{self organizing networks}
\acrodef{SVM}{support vector machine}
\acrodef{TCP}{transmission control protocol}
\acrodef{TL}{Transfer learning}
\acrodef{TDD}{time division duplex}
\acrodef{TD3}{Twin Delayed Deep Deterministic policy gradient}
\acrodef{TDMA}{time division multiple access}
\acrodef{TTI}{transmission time interval}
\acrodef{TX}{transmitter}
\acrodef{UDP}{user datagram protocol}
\acrodef{UE}{user equipment}
\acrodef{UL}{uplink}
\acrodef{V2X}{vehicle-to-everything}
\acrodef{WLAN}{wireless local area network}

\usepackage{acronym}
\usepackage{xspace}
\usepackage{verbatim}
\usepackage[usenames]{color}
\usepackage{array}
\usepackage{multicol}
\usepackage{algorithm}
\usepackage{algpseudocode}
\usepackage{amsmath,amsfonts,amssymb,amsthm}
\usepackage{graphicx}
\usepackage[tight,footnotesize]{subfigure}
\usepackage{float}
\usepackage{epstopdf}
\usepackage{cite}
\usepackage{url}

\usepackage[english]{babel}
\usepackage{enumitem} 
\usepackage[left=1.3cm,right=1.3cm,top=1.85cm,bottom=4.4cm]{geometry}
\usepackage[font=small,skip=2pt]{caption}
\usepackage[utf8]{inputenc}
\usepackage[mathscr]{euscript}
\usepackage{textcomp}
\usepackage{xcolor}
\usepackage{bbm}

\DeclareMathAlphabet\mathbfcal{OMS}{cmsy}{b}{n}

\newtheorem{Rem}{Remark}
\usepackage[english]{babel}
\usepackage{amsthm}

\newtheorem{problem}{Problem}

\DeclareRobustCommand\optionalsec[1]{%
  \ifnum\pdfstrcmp{#1}{\thesection}=0\else#1.\fi
}

\DeclareMathOperator*\argmax{arg \, max}		
\DeclareMathOperator*\argmin{arg \, min}		
\DeclareMathOperator*\maximize{max.}		
\DeclareMathOperator*\subject{subject \ to}

\newcommand{\field}[1]{\mathbb{#1}}

\newcommand{\set}[1]{\mathcal{#1}}

\newcommand{\pluseq}{\mathrel{+}\mathrel{\mkern-2mu}=}

\newcommand{\R}{{\field{R}}}   

\newcommand{\NN}{{\field{N}}}

\newcommand{\ma}[1]{\boldsymbol{\mathbf{#1}}} 
\newcommand{\ve}[1]{\boldsymbol{\mathbf{#1}}} 

\newcommand{\vx}{\ve{x}}
\newcommand{\vz}{\ve{z}}

\newcommand{\vq}{\ve{q}}

\newcommand{\vr}{\ve{r}}

\newcommand{\vv}{\ve{v}}

\newcommand{\vth}{\ve{\theta}}

\newcommand{\N}{{\set{N}}}

\newcommand{\C}{{\set{C}}}
\newcommand{\T}{{\set{T}}}

\newcommand{\Hh}{{\set{H}}}
\newcommand{\Ss}{{\set{S}}}

\newcommand{\X}{{\set{X}}}
\newcommand{\Y}{{\set{Y}}}

\newcommand{\Lag}{\set{L}}

\begin{document}    
\title{Fast and Scalable Network Slicing by Integrating Deep Learning with Lagrangian Methods}
\author{
\IEEEauthorblockN{Tianlun Hu\IEEEauthorrefmark{1}\IEEEauthorrefmark{4}, 
				  Qi Liao\IEEEauthorrefmark{1}, 
				  Qiang Liu\IEEEauthorrefmark{2},
                    Antonio Massaro\IEEEauthorrefmark{3}
				  and Georg Carle\IEEEauthorrefmark{4}}
\IEEEauthorblockA{ 
	\IEEEauthorrefmark{1}Nokia Bell Labs, Stuttgart, Germany\\
	\IEEEauthorrefmark{2}University of Nebraska Lincoln, United States\\
        \IEEEauthorrefmark{3}Nokia Bell Labs, Paris, France\\
	\IEEEauthorrefmark{4}Technical University of Munich, Germany\\
Email: \IEEEauthorrefmark{1}\IEEEauthorrefmark{4}\url{tianlun.hu@nokia.com}, 
	   \IEEEauthorrefmark{1}\url{qi.liao@nokia-bell-labs.com}, 
	   \IEEEauthorrefmark{2}\url{qiang.liu@unl.edu}, \\
          \IEEEauthorrefmark{3}\url{antonio.massaro@nokia-bell-labs.com}, 
	   \IEEEauthorrefmark{4}\url{carle@net.in.tum.de}}
	   \vspace{-0.4in}
}

\maketitle

\begin{abstract}
Network slicing is a key technique in 5G and beyond for efficiently supporting diverse services.
Many network slicing solutions rely on deep learning to manage complex and high-dimensional resource allocation problems. However, deep learning models suffer limited generalization and adaptability to dynamic slicing configurations. In this paper, we propose a novel framework that integrates constrained optimization methods and deep learning models, resulting in strong generalization and superior approximation capability. 
Based on the proposed framework, we design a new neural-assisted algorithm to allocate radio resources to slices to maximize the network utility under inter-slice resource constraints. The algorithm exhibits high scalability, accommodating varying numbers of slices and slice configurations with ease.
We implement the proposed solution in a system-level network simulator and evaluate its performance extensively by comparing it to state-of-the-art solutions including deep reinforcement learning approaches.
The numerical results show that our solution obtains near-optimal quality-of-service satisfaction and promising generalization performance under different network slicing scenarios.
\end{abstract}

\makeatletter{\renewcommand*{\@makefnmark}{}\footnotetext{This work was supported by the German Federal Ministry of Education and Research (BMBF) project 6G-ANNA.}\makeatother}
\makeatletter{\renewcommand*{\@makefnmark}{}\footnotetext{Qiang Liu\IEEEauthorrefmark{2}'s work is partially supported by the US National Science Foundation under Grant No. 2212050.}\makeatother}

\section{Introduction}\label{Section:Introduction}
Network slicing has been widely investigated in 5G and beyond to support different network services in terms of cost-efficiency, flexibility, and assurance~\cite{salvat2018overbooking}. 
The ever-increasingly disaggregated network elements with fine-grained controllability may lead to volatile network dynamics in various aspects, e.g., admission and departure of slices in small time scales~\cite{d2020sl}.
As a result, allocating radio resources to dynamic network slices becomes even more challenging.

The problem of resource allocation in network slicing has been extensively studied in the scenario of individual cells, where allocations are mostly optimized under the assumption that the resource demand of slices is known in advance. 
Existing works derive their solutions by formulating analytical closed-form models and solving the network slicing problem using constrained nonlinear optimization methods. 
In \cite{A-2023ResourceAI}, the authors initially formulated and streamlined the slice-wise resource allocation problem by finding the upper and lower bound of network utility using the Lagrangian method. 
Subsequently, a sub-optimal solution was obtained using a greedy algorithm. Although the derived simplified model is effective, it is still tailored to specific slice configurations. 
In \cite{F-Leconte2018ARA}, authors proposed a flexible slice deployment solution with dynamic slice configurations, which formulated a slice model with adjustable parameters and solved resource partition with an optimization process. 
However, recent findings~\cite{liu2021constraint, qiang2021onslicing} show that these approximated models cannot accurately represent diverse demand and performance of slices.

With recent advances in machine learning, \ac{RL} methods have been increasingly explored to tackle complex allocation problems in dynamic mobile networks. 
Zhou \emph{et al.} \cite{D-Zhou2021RANRS} designed a multi-agent RL framework based on Q-Learning to determine the optimal joint resource allocation by using a coordinated Q-table, which alleviates the inter-slice resource constraints. 
However, this solution cannot scale to large state and action spaces.
Our previous work \cite{Hu2022InterCellReDRL}
investigated coordinated multi-agent \ac{DRL} to handle the high-dimensional continuous action space and complex resource optimization in network slicing, where the inter-slice resource constraints are embedded in the designed architecture of neural networks. 
However, the proposed solution was explicitly trained for a fixed network scenario, and can hardly be generalized for different slice setups in terms of slice type and slice number.
Liu \emph{et al.} \cite{G-Liu2020DeepSlicingDR} introduced an approach to combine \ac{DRL} and optimization process for slicing resource allocation in a single cell scenario. 
Yet, it also lacks the discussion of generalizing the solution to different multi-cell network scenarios with flexible slice setups. 

In this paper, we present a novel algorithm, called \ac{IDLA}, that optimizes slicing resource allocation and can be generalized to adapt to arbitrary slice combinations under time-varying dynamics.
The main contributions of this work are listed as follows:
\begin{itemize}
    \item We propose a novel framework that integrates deep learning models (that have approximation capability) and constrained optimization methods (that have strong generalization) and can generalize to arbitrary slice combinations.
    \item First, we derive a general \ac{DNN} model to approximate the slice network utility, that is capable of handling slices under different requirements. 
    \item Then, by leveraging the efficient computation of the partial derivatives of the slice utility function approximated by the \ac{DNN} model, we design a Lagrangian method for resolving the optimal resource allocation on a per-slice basis, while adhering to inter-slice resource constraints. 
    \item We evaluate the proposed algorithm in a system-level network simulator, where numerical results show that our algorithm obtains near-optimal \ac{QoS} satisfaction and promising generalization performance as compared to the state-of-the-art solutions including the widely used \ac{DRL} approaches.
\end{itemize}
This paper is organized as follows. We define the system model in Section \ref{Section:SysModel} and formulate the slice-aware resource allocation problem in Section \ref{Section:Problem}. In Section \ref{Section:Solution} we propose the solutions with \ac{DNN}-based slice utility estimator and constrained nonlinear optimization. The numerical results are shown in Section \ref{Section:Results}. We conclude this paper in Section \ref{Section:Conclusion}.

\section{System Model}\label{Section:SysModel}
\begin{figure}
    \centering
    \includegraphics[width=0.38\textwidth]{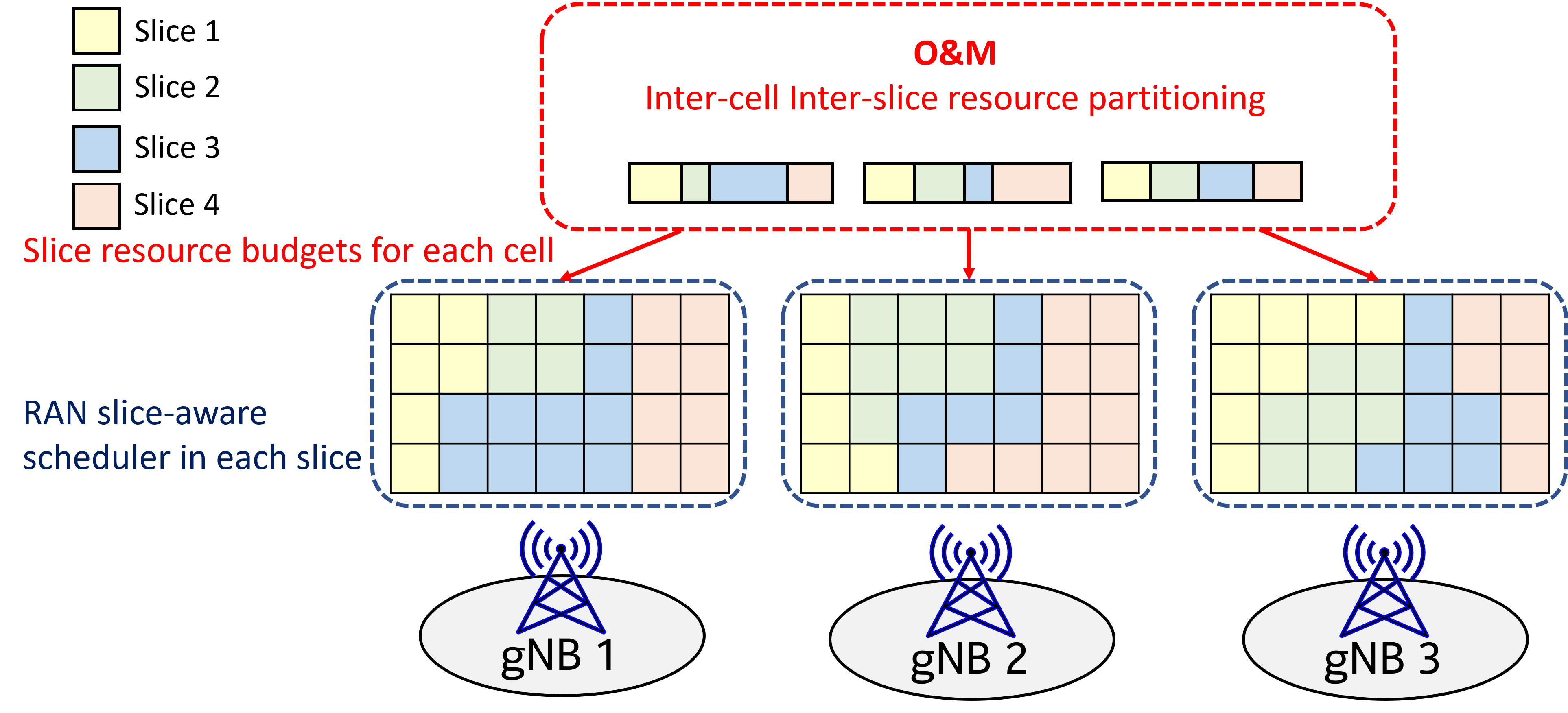}
    \vspace{0ex}
    \caption{Dynamic slicing resource partition}
    \label{fig:RANSlicing}
\end{figure}
We consider a discrete-time network system that comprises a set of cells denoted by $\C:=\{1, 2, ..., C\}$. The set of slices in each cell $c\in\C$ can be time-varying, denoted by $\Ss_c(t):=\{1, 2, ..., S_c(t)\}$, where $S_c(t)$ is the number of slices served by cell $c$ at time slot $t\in\NN_0$. Each slice $s\in\Ss_c(t)$ in cell $c$ needs to meet the predefined \ac{QoS} requirements, e.g., throughput and delay requirements denoted by $\phi_s^\ast$ and $d_s^\ast$ respectively. Note that although here the slices are defined by throughput and delay requirements, the problem formulation and the proposed approach in the following sections can be generalized to a broader set of requirements. 

As illustrated in Fig. \ref{fig:RANSlicing}, \ac{OAM} dynamically partitions the inter-slice resource to provide per-slice resource budgets to each cell periodically. Within each cell, the \ac{RAN} scheduler allocates \acp{PRB} to individual services, using the provided resource budgets as upper-bound constraints. The focus of this paper is to solve inter-slice resource partitioning problem in network \ac{OAM}, because we aim to develop a general slicing resource partitioning solution compatible with \ac{RAN} schedulers from different network providers. At each time slot $t$, \ac{OAM} optimizes slicing resource partitioning $\vx_c(t)$ for each cell $c$,  i.e., the ratio of the radio resource to allocate to each slice, given by
\begin{align}
    \vx_c(t)& :=\left[x_{c, 1} (t), \ldots, x_{c, S_c(t)}(t)\right] \in \X_c(t), ~\forall c\in\C, \label{eqn:action_def} \\
    \mbox{where } & \X_c(t) :=\left\{[0, 1]^{S_c(t)}\Big|\sum_{s\in\Ss_c(t)}x_{c,s}(t)\leq 1\right\}. \label{eqn:action_space}
\end{align}

Let $\vx(t):=[\vx_1(t), \ldots, \vx_C(t)]$ be a collection of the per-cell slicing partitioning, the performance of each slice $s\in\Ss_c(t)$ in cell $c\in\C$ at time $t$ is measured by the \ac{QoS} satisfaction level $r_{c,s}(\vx(t))$, defined as
\begin{equation}
r_{c,s}(\vx(t)):= \min\left\{\frac{\phi_{c,s}(\vx(t))}{\phi^\ast_s}, \frac{d^\ast_s}{d_{c,s}(\vx(t))}, 1\right\}, 
\label{eqn:local_utility}
\end{equation}
where $\phi_{c,s}(\vx(t))$ and $d_{c,s}(\vx(t))$ are the throughput and delay  associated with slice $s$ at cell $c$ at time slot $t$, respectively.
This performance metric takes the minimum between throughput and delay and is upper bounded by $1$, such that both requirements need to be met to achieve the satisfaction level of $1$.

\begin{Rem}
Theoretically, due to the inter-cell and possibly inter-slice interference, the achievable throughput and delay not only depends on the locally allocated resource to its own slice and cell, but also the resource occupation of other slices in the neighboring cells. Thus, in \eqref{eqn:local_utility}, the \ac{QoS} metric $r_{c,s}$ is written as a function of the global slicing partitioning $\vx(t)$.
\label{rem:interference}
\end{Rem}

\section{Problem Formulation}\label{Section:Problem}
Our objective is to find an efficient and scalable solution to optimize the utility of \ac{QoS} satisfaction over all slices and cells by optimizing slicing resource partitioning at each time slot. The per-slot optimization problem is formulated in Problem \ref{prob:Opt}.
\begin{problem}[Global Problem]
\label{prob:Opt}
\begin{equation}
    \begin{split}
        \maximize_{\vx(t)} ~ & U(\vr(\vx(t))) \\
        \subject ~& \vr(\vx(t)):=\big[r_{c,s}(\vx(t)): c\in\C, s\in\Ss_c(t)\big], \\
         & \eqref{eqn:action_def}, \eqref{eqn:action_space}, \eqref{eqn:local_utility}, ~ \forall t.
    \end{split}
\end{equation}
\end{problem}

Note that the utility function can be defined based on various system designs. For example, a common utility function to consider the fairness is the sum of the logarithmic function of the local performance metric:
\begin{align}
U(\vr(\vx(t)))&:=\sum_{s\in\Ss_c(t), c\in\C} \log\left(r_{c,s}(\vx(t)) + 1\right).\label{eqn:log_utility}
\end{align}

In this paper, we use \eqref{eqn:log_utility} as an example of the utility function, however, by leveraging the superior approximation capability of deep learning, our proposed approach can be applied to a wide range of utility functions. The challenge of solving Problem \ref{prob:Opt} is multifaceted. Firstly, the utility function's complexity poses a challenge to function approximation, particularly due to limited measurements in \ac{OAM}. In contrast to \ac{RAN}, where user and channel feedback can be collected with fine time granularity (e.g., in milliseconds), \ac{OAM} only collects averaged cell- and slice-level \acp{KPI} with a coarse granularity (e.g., in minutes). Consequently, deriving closed-form expressions becomes extremely challenging. Secondly, the flexible slice configurations and inter-slice constraints further complicate the problem, resulting in slow convergence and poor adaptability of deep learning-based approaches.  Finally, \ac{OAM}'s high scalability demand, e.g., up to over 100k cells, makes it challenging to use either large global deep learning models or collaborative multi-agent local models that require extensive exploration to learn from scratch.

\section{Proposed Solutions}\label{Section:Solution}
In this section, we propose \ac{IDLA} algorithm, to address the aforementioned challenges.
First, we design and train a \ac{DNN} to approximate the per-slice utility function. 
Then, with the derived slice-based utility model, we decompose Problem \ref{prob:Opt} into distributed cell-based resource allocation problems with inter-slice resource constraints.
This decomposition allows the \ac{IDLA} algorithm to adapt to a flexible number of slices per cell. 
Next, we use the Lagrangian method to solve the constrained decomposed problem, where the partial derivatives can be efficiently computed based on the \ac{DNN}-based utility model.
Finally, by leveraging the automatic differentiation engine of deep learning libraries, we improve the efficiency of the Lagrangian method.
\subsection{Slice-based \ac{QoS} Prediction using \ac{DNN}}\label{sec:Pred_training}
The complexity of the global utility function in \eqref{eqn:log_utility} is caused by the dependency of each local utility $U_{c,s}(r_{c,s}(\vx(t)))$ on the global slice resource partition $\vx(t)$. Our idea is to investigate whether each local per-slice \ac{QoS} satisfaction level $r_{c,s}(t)$ can be approximated by a single general \ac{DNN} $f_{\vth}(x_{c,s}(t), \vz_{c,s}(t)), \forall c, s$ based on the local observations only, including the allocated slice resource $x_{c,s}(t)$ and a set of slice-based \acp{KPI} $\vz_{c,s}(t)$, i.e., to find 
\begin{equation}
f_{\vth}(x_{c,s}(t), \vz_{c,s}(t))\approx r_{c,s}(\vx(t)), \forall s\in\Ss_c(t), c\in\C, 
\label{eqn:utility_approximator}
\end{equation}
where the \ac{DNN} is parameterized by $\vth$.

{\bf Data Collection:} The \ac{DNN} model serves as a general slice-based \ac{QoS} estimator, trained on slice-wise data collected from network \ac{KPI}s of different cells and slice configurations. Such data are collected in network \ac{OAM} periodically, e.g., every $15$ minutes, as a standard practical setting. Based on the experts' prior knowledge, to predict the slice utility $r_{c,s}(t)$, computed with the achievable throughput $\phi_{c,s}(t)$ and latency $d_{c,s}(t)$, $\forall c, s$, the following network \acp{KPI} are highly correlated:

\begin{itemize}
    \item Per-slice required throughput $\phi_s^{\ast}$ and required delay $d_s^{\ast}$;
    \item Per-slice \ac{PRB} utilization ratio $p_{c,s}(t)$, defined as the ratio of the \acp{PRB} occupied by the slice, which can be seen as the input of the allocated resource $x_{c,s}(t)$. This is because, if resource $x_{c,s}(t):=p_{c,s}(t)$ was allocated to the slice, the corresponding achieved throughput and delay would be the same as $\phi_{c,s}(t)$ and $d_{c,s}(t)$, respectively;
    \item The previous $H$ states of per-slice average number of active users $\vv_{c,s}^{(H)}(t):=\left[v_{c,s}(t-H), \ldots, v_{c,s}(t-1)\right]$;
    \item The previous $H$ states of per-slice average \ac{CQI} $\vq_{c,s}^{(H)}(t):=[q_{c,s}(t-H), \ldots, q_{c,s}(t-1)]$.
\end{itemize}

Note that we collect multiple historical states of the average number of active users and \ac{CQI}, in the hope that the historical slice states not only capture temporal correlation, but also reflect some hidden information extracted from the missing global states, e.g., experienced inter-cell and inter-slice interference. Also, we follow the realistic assumption that for model inference, the real-time $v_{c,s}(t)$ and $q_{c,s}(t)$ are unknown while only the previous states within $[t-H, t-1]$ are available.

Thus, a set of the local observations as the input samples during a time period $[1, T]$ is then denoted by:
\begin{align}
\X_{t=1}^T:= &\{(x_{c,s}(t), \vz_{c,s}(t)): \mbox{ for } t= 1, \ldots, T, \forall c, s\}, \label{eqn:sample_in}\\ 
\mbox{where } & \vz_{c,s}(t):=[\vv_{c,s}(t), \vq_{c,s}(t), \phi_s^{\ast}, d_s^{\ast}]\in\R^{2H+2}, \label{eqn:sample_z}
\end{align}
while the set of output samples is denoted by:
\begin{equation}
\Y_{t=1}^T:=\left\{r_{c,s}(t):\mbox{ for }t= 1, \ldots, T, \forall c, s\right\},
\label{eqn:sample_out}
\end{equation}
where the \ac{QoS} satisfaction level $r_{c,s}(t)$ is computed by \eqref{eqn:local_utility} based on the observed throughput $\phi_{c,s}(t)$ and delay $d_{c,s}(t)$.

{\bf Local Utility Approximation:} 
We learn a general slice \ac{QoS} estimator $f_{\vth}:\R^{2H+3}\to\R: (x_{c,s}, \vz_{c,s}) \mapsto r_{c,s}$, as defined in \eqref{eqn:utility_approximator}, such that the local utility in \eqref{eqn:log_utility} can be approximated  by:
\begin{equation}
    U_{c,s}(r_{c,s}(\vx(t))) \approx \log(f_{\vth}(x_{c,s}(t), \vz_{c,s}(t)) + 1).
    \label{eqn:QoS_estimator}
\end{equation}

With the collected data \eqref{eqn:sample_in} and \eqref{eqn:sample_out}, we can train a \ac{MLP} with $[x_{c,s}(t), \vz_{c,s}(t)]\in\X_{t=1}^T$ as inputs and $r_{c,s}(t)\in \Y_{t=1}^T$ as output. Because $f_{\vth}(\cdot)$ is a general distributed model that can apply to any slice, at each time $t$ the samples from any cell and slice can contribute to model training, leading to a much higher sample efficiency and faster learning speed than training a large number of local models for distinct cells and slices. Moreover, a general model, that includes the throughput and delay requirements into the input features, can handle flexible slice configurations, even with unseen requirements. 
\begin{Rem}
More details of data augmentation for unseen samples of different slice configurations will be given in Section \ref{result:estimator_training}. Moreover, in Section \ref{result:estimator_training} we validate the viability of learning \eqref{eqn:utility_approximator} not only on the simulated data, but also on a collected real dataset from a commercial LTE network.
\end{Rem}
\subsection{Lagrangian Method for Slicing Resource Partitioning}\label{sec:DNN_optimize}
With \ac{DNN}-based utility approximation \eqref{eqn:QoS_estimator} at hand, we can decompose Problem \ref{prob:Opt} into independent per-cell optimization problems with intra-cell and inter-slice resource constraints. For each cell $c\in\C$ at time $t$, optimization problem is written as:
\begin{problem}[Decomposed Local Problem]
\label{prob:local_Opt}
\begin{equation*}
         \maximize_{\vx_c} ~ F(\vx_c) := \sum_{s\in\Ss_c(t)}  \log\Big(f_{\vth}\big(x_{c,s}(t), \hat{\vz}_{c,s}(t)\big) + 1\Big) 
\end{equation*}
        \begin{equation}
         \subject  ~ \eqref{eqn:action_def}, \eqref{eqn:action_space}, ~ \forall t, \forall c\in\C, \label{eqn:DNN_optimization}
\end{equation}
where $\hat{\vz}_{c,s}(t)$ is the local observations defined in \eqref{eqn:sample_z}.
\end{problem}

Problem \ref{prob:local_Opt} is a classical constrained non-linear optimization problem.  
Note that the objective in \eqref{eqn:DNN_optimization} is a monotonic non-decreasing function over $\vx\in\R_+$, i.e., we have $F(\vx_c')\geq F(\vx_c)$ if $\vx_c'\geq \vx_c$ (entrywise greater). Therefore, the optimal solution to the problem with the equality constraint is also an optimal solution to the original problem, and we can solve it by using the Lagrange multiplier method. Since the problem is independently formulated for each time slot $t$ and cell $c\in\C$, hereafter in this subsection we omit the index of $t$ for brevity.

For each cell $c\in\C$, the Lagrangian is given by:
\begin{equation}
        \Lag(\vx_c, \lambda_c) := \sum_{s\in\Ss_c} \log f_{\vth}(x_{c, s}) + 
        \lambda_c \big(1-\sum_{s\in\Ss_c} x_{c, s}\big),
    \label{eqn:Lagrangian}
\end{equation}
where $f_{\vth}(x_{c, s}):=f_{\vth}\big(x_{c,s}(t), \hat{\vz}_{c,s}(t)\big)$ is the learned \ac{DNN} in \eqref{eqn:DNN_optimization}, and $\lambda_c \in \R^+$ is the real non-negative Lagrangian multiplier. Then, we can solve the primal and dual problems:
\begin{align}
\vx_c^{\ast}(\lambda_c) &= \argmax_{\vx_c\in\R^+}\Lag(\vx_c, \lambda_c), \\
\lambda_c^{\ast} &=\argmin_{\lambda_c\geq 0 }\Lag(\vx_c^{\ast}(\lambda_c), \lambda_c),
\end{align}
by computing the partial derivatives with respect to each variable and performing \ac{GD} iteratively: 
\begin{equation}
\small
\left\{
\begin{aligned} 
x_{c,s}^{(i+1)} := & \left[x_{c,s}^{(i)} + \delta_x^{(i)} \cdot \frac{\partial\Lag_c\left(\vx_c^{(i)}, \lambda_c^{(i)}\right)}{\partial x_{c,s}^{(i)}}\right]_{+}, \forall s\in \Ss_c \\
\lambda_c^{(i+1)} := &\left[\lambda_c^{(i)} - \delta_{\lambda}^{(i)} \cdot \left(1 - \sum_{s\in\Ss_c} x_{c,s}^{(i+1)}\right)\right]_{+},
\end{aligned}
\right.
\label{eqn:GD_iteration}
\end{equation}
where $i$ is the index of iteration, $\delta_x$ and $\delta_{\lambda}$ are the positive updating rates of $x_{c,s}, \forall s\in\Ss_c$ and $\lambda_c$, respectively, and $[x]_{+}$ is equivalent to $\max\{x, 0\}$. The partial derivative of $\Lag_c$ with respect to $x_{c,s}$, $\forall s\in\Ss_c$ is given by:
\begin{align}
\small
    \frac{\partial \Lag_c\left(\vx_c^{(i)}, \lambda_c^{(i)}\right)}{\partial x_{c,s}^{(i)}} = \frac{1}{f_{\vth}\left(x_{c,s}^{(i)}\right)+1} \cdot \frac{\partial f_\theta\left(x^{(i)}_{c,s}\right)}{\partial x^{(i)}_{c,s}} - \lambda_c^{(i)}.
    \label{eqn:DNN_derivative}
\end{align}

\subsection{Efficient Implementation to Improve the Performance of Lagrange Multiplier Method}\label{sec:Lagrangian_method}
One major limitation of the Lagrangian methods is that, if the function is non-linear and non-convex, there might exist multiple solutions or folds on the functional surface, and searching on one path may easily get stuck in a local optima. 
To overcome this, we exploit the automatic differentiation engine of deep learning libraries, and design a robust search strategy. 

By using the automatic differentiation module \emph{torch.autograd} of PyTorch \cite{paszke2017automatic}, we can efficiently compute the partial derivative of the trained function with respect to any input variables on tensors, e.g., the partial derivative $\partial f_\theta\left(x^{(i)}_{c,s}\right)/\partial x^{(i)}_{c,s}$ in \eqref{eqn:DNN_derivative}. This allows fast parallel computing of multiple searching paths. Thus, we propose the following search strategy:

\begin{itemize}
    \item [(1)] Based on the assumption that the network states between two successive time steps change smoothly, we propose to initialize the starting points for the optimization of each time slot $t$ with the optimized solution of the previous time slot $t-1$, i.e., $\vx_{c}^{(0)}(t) := \vx_{c}^*(t-1)$;
    \item [(2)] To find a better (possibly local) optima, we take $P$ neighboring points near $\vx_{c}^{(0)}(t)$ and run the \ac{GD} optimizations from all $P$ initial points in parallel. After \ac{GD} optimizations have finished, we select the best solution among them.
\end{itemize}

The proposed \ac{IDLA} algorithm is summarized in Algorithm \ref{algo:DNN_optimization}, where $\N(\ve{\mu}, \ma{\Sigma})$, $i^{(\max)}$, and $\eta$ denote the normal distribution for taking neighboring points with mean $\ve{\mu}$ and covariance matrix $\ma{\Sigma}$, the maximum iteration steps, and criterion for stopping iteration, respectively.

\begin{algorithm}
    \caption{\ac{IDLA} Algorithm}
    \label{algo:DNN_optimization}
    \begin{algorithmic}[1]
        \For{$t\in \T$ and $c\in\C$}
            \State $i \leftarrow 0$
            \State 
                $\vx_c^{(i)}(t)\leftarrow\left\{
                \begin{aligned}
                &\text{\emph{default action}} , & \mbox{if } t = 0  \\
                &\vx_{c}^*(t-1) , & \text{Otherwise}
                \end{aligned}\right.$
            \State Take $P$ neighboring points as:
            \State $\vx_{c_{p}}^{(i)}(t):=\vx_c^{(i)}(t) + \ve{\epsilon}$, $p\in[1, ..., P]$ with $\ve{\epsilon} \in \N(\ve{\mu}, \ma{\Sigma})$
            \State {\bf Parallelly compute} for all $p\in[1, ..., P]$:
            \State Initialize Lagrangian multiplier $\lambda_{c_p}^{(i)}$
            \State Initialize update rate $\delta_{x_p}^{(i)} >0, \delta_{\lambda_p}^{(i)} >0$
            \While{$i\leq i^{(\max)}$ and $\|\vx_{c_p}^{(i)}(t)-\vx_{c_p}^{(i-1)}(t)\|\geq \eta$}
                    \State Compute partial derivative with \eqref{eqn:DNN_derivative}  $\forall s\in\Ss_c(t)$ 
                    \State Update optimization variables and Lagrangian multipliers with \eqref{eqn:GD_iteration}
                \State Decrease update rate $\delta_{x_p}^i, \delta_{\lambda_p}^i$
                \State $i \pluseq 1$
            \EndWhile
            \State $\vx_{c_p}^{\ast}(t)\leftarrow \vx_{c_p}^{(i)}(t)$
            \State Choose the best solution among all $P$ points that provides the highest utility:
            \State $\vx_{c}^*(t) := \argmax_{\vx_{c_{p}}^{\ast}(t)} \sum_{s\in\Ss_c(t)} \log\left(f_{\vth}\left(x_{c_{p}, s}^{\ast}(t)\right)\right)$.
        \EndFor
    \end{algorithmic}
\end{algorithm}

\section{Performance Evaluation}\label{Section:Results}
In this section, we evaluate the performance of the proposed algorithm by implementing it in a system-level network simulator \cite{SeasonII}, which can imitate real network systems well with configurable user mobility, and slicing services. 
We compare the real-time processing performance of the \ac{IDLA} scheme with two state-of-art schemes including a cell-wise \ac{DRL} scheme and a traffic-aware baseline that allocates resources proportionally to data traffic demand per slice. We also compare it with an oracle scheme obtained by brute force optimization with the theoretically optimal performance. In addition, we explore the flexibility of the \ac{IDLA} algorithm when facing slice configuration changes and its transferability from sample-collecting network configurations to a new network configuration.

\subsection{Network Setting}\label{result:Network_setup}
\begin{figure}
    \centering
    \includegraphics[width=0.42\textwidth, height=1.7in]{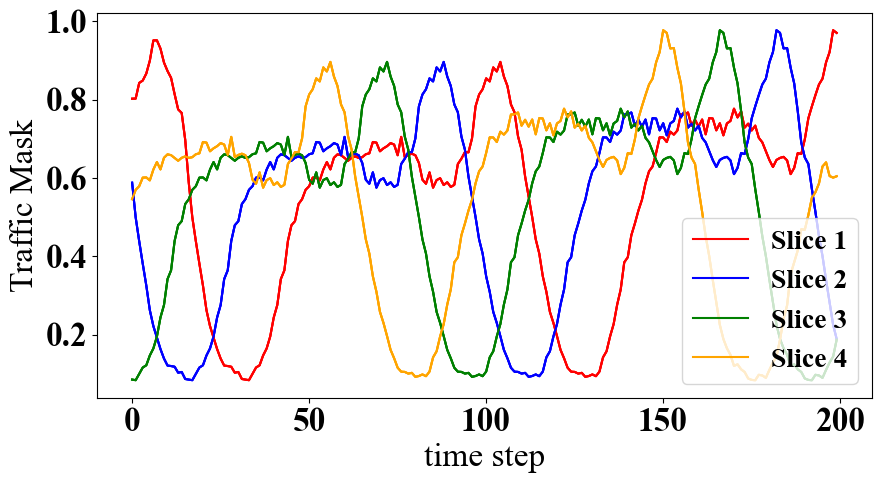}
    \caption{Traffic mask to imitate the dynamic slice traffic}
    \label{Fig:TrafficMask}
\end{figure}

We built a network system consisting of $4$ three-sector base stations with the operating frequency band of $2.6$ GHz, i.e., $C=12$ cells. We defined $4$ types of services, where the slice combination $\Ss_c(t)$ can be configurable and time-varying. Each service has different requirement as average user throughput $\phi_s^*$ for $s=1,2,3,4$ defined as $\{2, 1, 1.5, 0.5\}$ MBit/s, respectively.
respectively. All cells are provided with the same bandwidth of $20$ MHz. In addition, to imitate the real user traffic, we apply a varying traffic mask $\tau_s(t)\in[0, 1]$, which is collected from a real network system, for each slice $s\in\Ss_c(t)$ to reflect the daily periodic pattern of per-slice user traffic. In Fig. \ref{Fig:TrafficMask} we present the first $200$ steps of the traffic mask. In the experiments, each step corresponds to $15$ minutes in real time.

\subsection{Sample Collection and Network \ac{QoS} Estimator Training}\label{result:estimator_training}
Before the training process of network \ac{QoS} estimator $f_{\vth}(\cdot)$, we collected the training samples from the built network scenario in the simulator following the pipeline introduced in section \ref{sec:Pred_training}. 

{\bf Data Augmentation.} We implemented a data augmentation strategy to cover a wider range of (unseen) sample space. The data augmentation strategy is summarized as follows:
\begin{itemize}
\item[(1)] For the per-slice samples that achieve lower network \ac{QoS} than the requirements, i.e., for $r_{c,s}(t)<1$, we generated augmented samples by replacing the \ac{QoS} requirements $(\phi_s^{\ast}, d_s^{\ast})$ in the input training sample with the achieved \ac{QoS} $(\phi_{c,s}(t), d_{c,s}(t))$ and replacing the \ac{QoS} satisfaction $r_{c,s}(t)$ (sample output) with $1$. Because, if the achieved $(\phi_{c,s}(t), d_{c,s}(t))$ were given as requirements, then the requirement would be met.
\item[(2)] Conversely, for the per-slice samples that achieve no less network \ac{QoS} than the requirement, i.e., for $r_{c,s}(t)=1$, we generated augmented samples by replacing the slice resource partition $x_{c,s}(t)$ in the input training samples with a random value $x'_{c,s}(t)\in[x_{c,s}(t), 1]$. Because the \ac{QoS} is upper bounded by $1$ based on \eqref{eqn:local_utility}, if more resource was given to the slice, due to the monotonicity of $r_{c,s}$ over $x_{c,s}$, the achieved $r_{c,s}$ would be $1$ as well.
\end{itemize}

{\bf Model training.} Then, for the training of \ac{QoS} estimator $f_{\vth}(\cdot)$, we built the \ac{DNN} model with \ac{MLP} architecture consisting of 4 hidden layers with the number of neurons $(36, 24, 16, 16)$. As proposed in \ref{sec:Pred_training}, to better capture the temporal correlation, we used $h=5$ steps of the historical network reports for estimator training, i.e., the training input $[x_{c,s}, \vz_{c,s}]\in\R^{13}$. The training data was collected from the network environment defined in \ref{result:Network_setup}. The model was trained for $200$ epochs on $75\%$ training samples and $25\%$ testing samples with Adam optimizer with respect to \ac{MAE} loss.

\begin{figure}
    \centering
    \includegraphics[width=0.42\textwidth, height=1.7in]{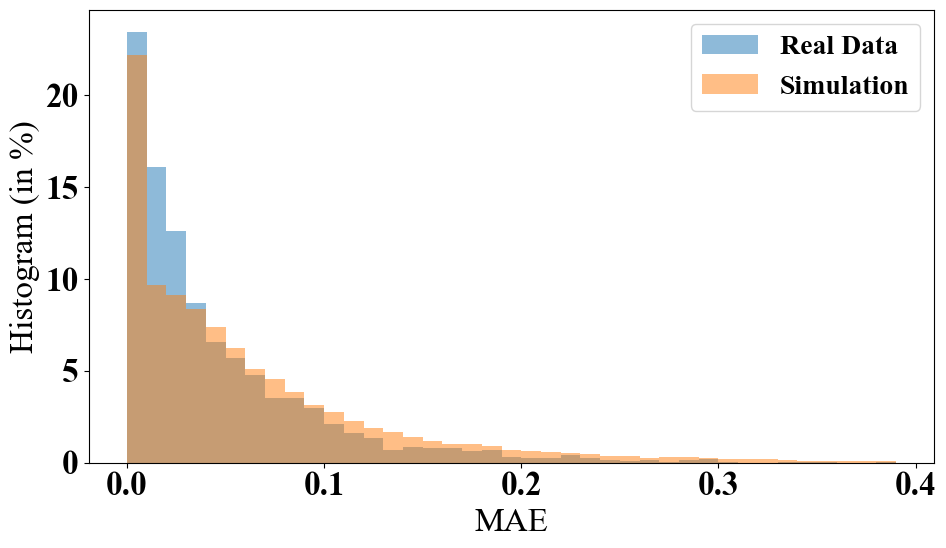}
    \caption{Network \ac{QoS} estimator \ac{MAE} histogram}
    \label{Fig:MAE_estimator}
\end{figure}

Fig. \ref{Fig:MAE_estimator} shows the histogram of the \ac{MAE} of network \ac{QoS} estimator after training was completed. To validate the viability of training a utility estimator, We investigated the \ac{DNN} model not only on the simulated data, but also on a dataset collected from a real commercial LTE network. By incorporating historical network reports, the estimator can provide accurate network \ac{QoS} predictions based on the given slice resource partition. The average MAE was $0.0639$ and $0.0573$ for the simulation and real dataset respectively. The close performances indicate that our method is valid for handling real network systems.

\begin{figure*}[!t]
    \captionsetup{justification=centering}
    \centering
    \includegraphics[width=0.9\textwidth]{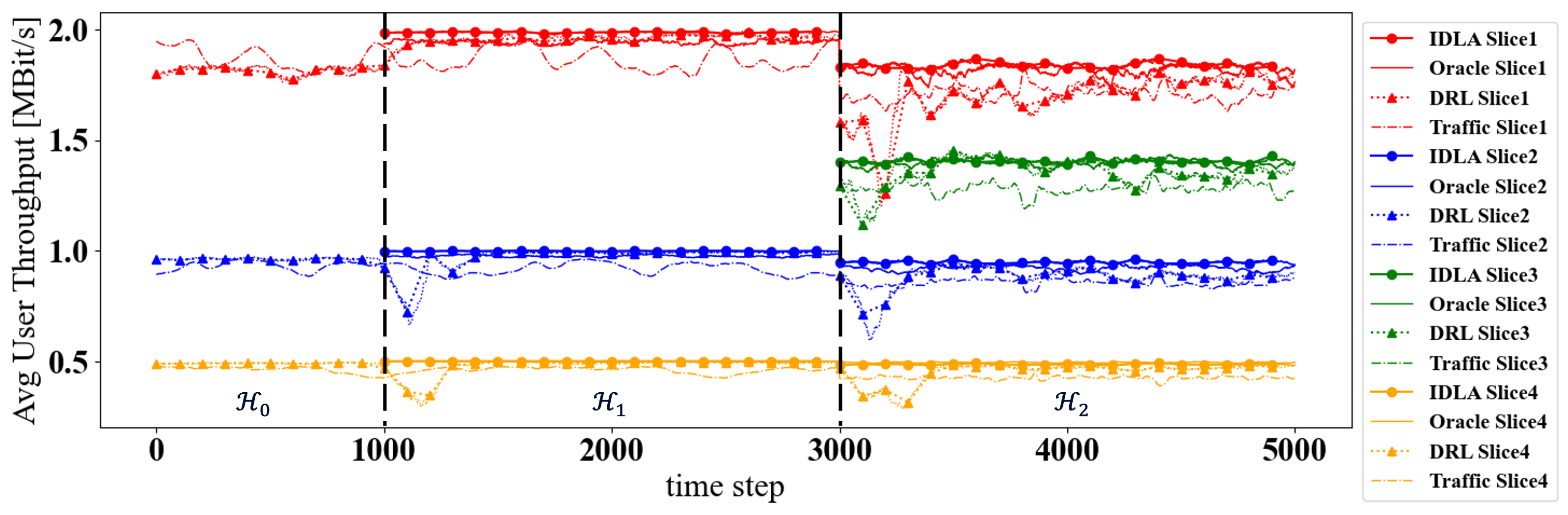}
    \caption{Comparison of average user throughput among schemes}
    \label{Fig:Compare_1}
\end{figure*}

\subsection{Performance Comparison}\label{result:Comparison}
With the derived slice \ac{QoS} estimator $f_{\vth}(\cdot)$ in hand, we further implement the optimization proposed in \ref{sec:Lagrangian_method} to obtain optimal resource partitions. We compare the performance of the following schemes:
\begin{itemize}
    \item \textbf{IDLA} scheme: our proposed algorithm in Algorithm~\ref{algo:DNN_optimization} with $P=5$ neighboring start points, where the offset $\epsilon$ for each point follows a normal distribution $\N(0, 0.05)$. 
    \item \textbf{\ac{DRL}} scheme: a distributed \ac{TD3} algorithm-based \ac{DRL} approach similar to our previous work \cite{Hu2022InterCellReDRL}, which solves cell-wise optimal slicing resource partitions regarding the reward defined by minimum of network \ac{QoS} \eqref{eqn:QoS_estimator} among all slices.
    \item \textbf{Traffic} scheme: a traffic-aware baseline that dynamically adapts slicing resource partitions in each cell proportionally to the current per-slice traffic amount, assuming perfect knowledge of traffic amount.
    \item \textbf{Oracle} scheme: an oracle scheme that provides the near-optimum for the constrained optimization problem. It is derived by using brute-force search for the optimal utility based on the pretrained $f_{\vth}(\cdot)$ over all potential combinations of $\vx_c\in\X_c$ with an interval of $0.05$.
\end{itemize}

To evaluate the performance of \ac{IDLA} against the other schemes, we implemented an online experiment in the network simulator with dynamic slice configuration, i.e., during the processing of the schemes, we changed the combination of network slices. For a fair comparison, we divide the whole online process into $3$ time periods, denoted by $\Hh_0$, $\Hh_1$, and $\Hh_2$ respectively:

\begin{itemize}
    \item $\Hh_0$ ($t\in[0, 1000)$): First, we set the network system with $3$ slices with combination $\Ss_c(t) := [1, 2, 4], t\in\Hh_0, c\in\C$. Both \textbf{IDLA} and \textbf{Oracle} schemes are under the stage of sample collection, while \textbf{DRL} is under the exploration phase for buffer collection without agent training. The \textbf{Traffic} scheme provides slice resource partitioning proportional to instantaneous slice traffic demands.
    \item $\Hh_1$ ($t\in[1000, 3000)$): The network keeps the same slice configuration as $\Hh_0$. The \textbf{IDLA} and \textbf{Oracle} schemes optimize resource allocation based on the pre-trained $f_{\vth}(\cdot)$ over the samples collected within $\Hh_0$, and \textbf{DRL} enter the phase of online training, with the samples collected within $\Hh_0$ also stored in the replay buffer.
    \item $\Hh_2$ ($t\in[3000, 5000]$): At $t=3000$, the network slice configuration changes to slice combination $\Ss_c(t):= [1, 2, 3, 4], t\in\Hh_2, c\in\C$, i.e., we introduce a new slice with the corresponding user group into the network system with the same resource constraints.
\end{itemize}

In Fig. \ref{Fig:Compare_1}, we compare the averaged per-slice user throughput $\phi_{c,s}(t)$ referring to its requirements $\phi_s^{\ast}$ over all cells for $s\in\Ss_c(t)$ of all schemes during the entire process $\{\Hh_0, \Hh_1, \Hh_2\}$. Note that in $\Hh_0$ and $\Hh_1$, there are only $3$ slices with index $[1, 2, 4]$, while later in period $\Hh_2$, we add a new slice with index $3$.
After the offline training of \ac{QoS} estimator $f_{\vth}(\cdot)$, the \textbf{\ac{IDLA}} provides the best performance among all schemes and faster convergence than \textbf{\ac{DRL}} scheme in both online processing phases $\Hh_1$ and $\Hh_2$. Moreover, \textbf{\ac{IDLA}} quickly adapts to the new slice configuration (with an added slice) in $\Hh_2$ and provides robust performance. On the contrary, due to the poor scalability of the cell-wise agent, the \textbf{DRL} scheme needs to retrain the model when the slice configuration changes. It is worth noting that since the total network resource remains the same after adding a new slice, the user throughputs in other slices decrease correspondingly to serve the users in the new slice.

\begin{figure}
    \centering
    \includegraphics[width=0.48\textwidth, height=1.6in]{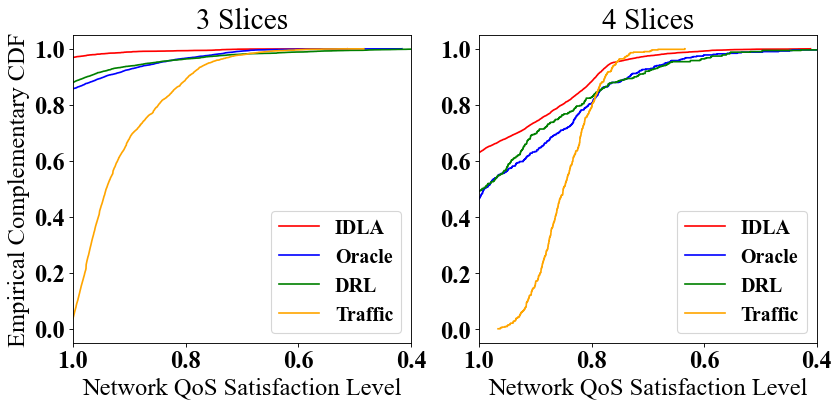}
    \caption{Comparison of network utility}
    \label{Fig:Compare_utility}
\end{figure}

To compare the converged performance of all schemes, in Fig. \ref{Fig:Compare_utility} we compare the empirical \ac{CDF} of the converged network \ac{QoS} satisfaction level of all schemes under both slice configurations. The \textbf{IDLA} scheme provides the highest probability of \ac{QoS} satisfaction under both slice configurations with $0.973$ and $0.629$ respectively, while \textbf{Oracle} and \textbf{DRL} schemes achieved similar converged \ac{QoS} satisfaction. Note that, theoretically, with brute-force search \textbf{Oracle} scheme should find a near-optimal solution if the utility estimator can be learned with $100\%$ accuracy. However, in this experiment, the performance of \textbf{Oracle} is not as good as \textbf{IDLA}, due to the estimation error of the utility estimator and the discretization of the searching grid space. 
In general, \textbf{IDLA} provides the best performance in terms of convergence rate, converged performance, and scalability.

\section{Conclusion}\label{Section:Conclusion}
\vspace{-0.05in}
In this paper, we propose a novel framework to integrate the strong generalization of the Lagrangian method and the superior approximation capability of the deep learning.
We developed \ac{IDLA} algorithm to solve the resource partitioning problem in network slicing with assured inter-slice resource constraint.
The results show that our proposed approach can provide near-optimal performance with fast convergence and high generality in comparison with state-of-the-art solutions.
In addition, we show the scalability of the proposed methods by deploying the derived model in different network scenarios with varying slicing configurations. 
With the slice-wise resource scheduler, our proposed algorithm provides high scalability and generality for fast and efficient deployment in real network systems.
\vspace{-0.05in}

\bibliographystyle{IEEEtran}
\bibliography{references}

\end{document}